Phase diagram and oxygen annealing effect of FeTe$_{1-x}$Se$_x$ iron-based superconductor

Yasuna Kawasaki[1,2,3], Keita Deguchi[1,2,3], Satoshi Demura[1,2,3], Tohru Watanabe[1,2], Hiroyuki Okazaki[1,3], Toshinori Ozaki[1,3], Takahide Yamaguchi[1,3], Hiroyuki Takeya[1,3] and Yoshihiko Takano[1,2,3]

[1] National Institute for Materials Science, 1-2-1 Sengen, Tsukuba 305-0047, Japan

[2] University of Tsukuba, 1-1-1 Tennoudai, Tsukuba, Ibaraki 305-0006, Japan

[3] JST-TRiP, 1-2-1 Sengen, Tsukuba 305-0047, Japan

Corresponding author: Yasuna Kawasaki

E-mail: keep_name@yahoo.co.jp

Phone: +81-29-851-2842

Fax: +81-29-859-2601

Adress: National Institute for Materials Science, 1-2-1 Sengen, Tsukuba 305-0047, Japan

Abstract

Phase diagrams of as-grown and O$_2$-annealed FeTe$_{1-x}$Se$_x$ determined from magnetic susceptibility measurement were obtained.  For as-grown samples, the antiferromagnetic order was fully suppressed in the range region $x \geq 0.15$, and weak superconductivity appeared when $x \geq 0.1$. Beginning at $x = 0.5$, weak superconductivity was found to evolve into bulk superconductivity. Interestingly, for O$_2$-annealed samples, complete suppression of magnetic order and the occurrence of bulk superconductivity were observed when $x \geq 0.1$.   We found that O$_2$-annealing induces bulk superconductivity for FeTe$_{1-x}$Se$_x$.   Oxygen probably plays a key role in the suppression of the magnetic order and the appearance of bulk superconductivity.

1. Introduction

In 2008, the new superconductor, LaFeAsO$_{1-x}$F$_x$, was discovered by Kamihara *et al* [1]. The discovery was quite surprising because iron compounds had been believed to be magnetic and not superconducting. This material consists of an alternate stacking of blocking and superconducting layers like the cuprates. Following this discovery, related materials have been found [2, 3]. Hsu *et al*. reported that anti-PbO-type FeSe shows superconducting transition with $T_c$ ~ 10 K [4]. This system, the so-called 11 type, is composed of only superconducting layers and has the simplest crystal structure among iron-based superconductors. Therefore, it is suitable for clarifying the mechanism of iron-based superconductivity.

In iron-based compounds, superconductivity is closely related to antiferromagnetism (AFM) since superconductivity appears as AFM order is suppressed by changing the carrier density. FeTe, which has a crystal structure similar that of FeSe, undergoes structural and magnetic transition at around $T_N$ = 70 K and does not show superconductivity [5 - 7]. However, the magnetic order of FeTe is suppressed by substituting Se for Te. The partial substitution of Se for Te induces a superconducting transition without changing the carrier density, which is consistent with the fact that Se and Te have the same valence.

In addition, this substitution changes the nesting vector $\mathbf{Q}_d$ = (0.5, 0) to $\mathbf{Q}_s$ = (0.5, 0.5) [8, 9]. The former gives rise to the double stripe collinear AFM state of FeTe while the latter corresponds to a single stripe collinear AFM order similar to that of FeAs-based superconductors. In fact, it is known that the antiferromagnetic wave vector $\mathbf{Q}_s$ correlates with superconductivity [10 - 12], which indicates that iron chalcogenides and iron pnictides have similar mechanisms for superconductivity. For FeTe, the emergence of a Fermi surface nesting associated with $\mathbf{Q}_d$ could be induced by excess Fe which supply a substantial amount of electrons [13], although this doping effect was not observed in angle-resolved photoemission spectroscopy (ARPES) [14, 15]. An excess Fe randomly occupies the Fe(2) site between the square planar sheets of Fe [6, 7]. As mentioned above, spin fluctuation and superconductivity are closely related for iron-based superconductors. As attempt to understand the mechanism of superconductivity for the 11 type, some phase diagrams represented by transition temperature as a function of Se or Te concentration has already been reported [16 - 19]. However, there are significant differences in the region of bulk or weak superconductivity and AFM order. Therefore, it is important to perform a systematic research and establish accurate phase diagrams.

We take particular note of a character of FeTe$_{1-x}$S$_x$ [20 - 23]. Although as-grown FeTe$_{1-x}$S$_x$ synthesized using the solid-state reaction method does not show superconductivity, the material exhibits superconducting transition with $T_c$ ~ 8 K by O$_2$-annealing. We expect that a similar effect occurs for O$_2$-annealed FeTe$_{1-x}$Se$_x$, since both Se and S are categorized as chalcogens.

In this paper, we report on the phase diagrams of as-grown and O$_2$-annealed FeTe$_{1-x}$Se$_x$ (0 ≤

$x \leq 0.5$) single crystals. We find that $O_2$-annealing induce bulk superconductivity.

2. Experimental Methods

Single crystals of FeTe$_{1-x}$Se$_x$ ($0 \leq x \leq 0.5$) were synthesized by the self-flux method. The powder of Fe (99.9%), the grains of Te (99.999%) and Se (99.999%) were sealed into an evacuated quartz tube with a nominal composition of FeTe$_{1-x}$Se$_x$. The quartz tube was sealed into another large-sized evacuated quartz tube since the small-sized quartz tube often cracked during cooling. The doubly sealed quartz tube was heated at 1100 °C for 20 h and then cooled down to 650 °C at the rate of 4 –5 °C/h. After furnace cooling, $O_2$-annealing was performed by sealing the quartz tube filled with $O_2$ gas of atmospheric pressure. The lattice constants $a$ and $c$ for the obtained crystals were estimated by x-ray diffraction using the 2θ – θ method with Cu-Kα radiation. The x-ray diffraction profile from 2θ = 10 to 70 deg. was collected using Mini Flex (Rigaku). The temperature dependence of magnetic susceptibility for the samples after both zero-field cooling (ZFC) and field cooling (FC) mode was measured using a SQUID magnetometer under an applied magnetic field of 10 Oe. The magnetic field was applied vertically to the $c$-axis of the FeTe$_{1-x}$Se$_x$ single crystal samples.

3. Results and discussion

Fig. 1. shows the lattice constants $a$ and $c$ estimated from the data of XRD as a function of Se concentration $x$. The values of $a$ and $c$ of as-grown and $O_2$-annealed samples were almost the same, and decreased linearly with increasing Se concentration, suggesting that a systematic substitution of Te by Se was successfully performed. The obtained values of each axis are in agreement with the previous report [24]. Furthermore, this result implies that $O_2$-annealing does not degrade the crystal structure of FeTe$_{1-x}$Se$_x$.

Fig. 2 (a). and (b). show the temperature dependence of magnetic susceptibility for the $O_2$-annealed FeTe$_{1-x}$Se$_x$ with $x = 0.1$ and 0.4 under several conditions. The shielding volume fraction was significantly enhanced with the increase of the $O_2$-annealing temperature. However, annealing temperatures exceeding 400 °C and long-time annealing exceeding 10 h were found to degrade the superconducting properties. The maximum shielding volume fraction and the highest $T_c$ were observed at 300 °C for 2 h in $O_2$ for both Se concentrations. We found that $O_2$-annealing effectively induces bulk superconductivity for FeTe$_{1-x}$Se$_x$. We therefore decided to anneal all samples under the condition.

Fig. 3 (a). - (d). show temperature dependence of magnetic susceptibility for as-grown and $O_2$-annealed samples with various Se concentrations. In addition, the Se concentration dependence of shielding volume fraction is plotted in Fig. 4. based on the results of magnetic susceptibility measurements. As shown in Fig. 3. and 4, for the as-grown samples, the long-range magnetic order was suppressed with increasing Se concentration and completely disappeared at $x = 0.15$. Although weak superconductivity was observed above $x = 0.15$, bulk superconductivity did not appear until $x = 0.5$. On the other hand, for the $O_2$-annealed samples, we found that only a 10 % substitution of Te by Se completely suppressed the magnetic order and induced bulk superconductivity. We summarize the accurate phase diagrams based on the magnetic susceptibility measurement for the as-grown and $O_2$-annealed samples, as shown in Fig. 5 (a). and (b). The as-grown samples showed the long-range AFM in the range $x < 0.15$ and weak superconductivity in $0.1 \leq x \leq 0.4$. Only the $FeTe_{0.5}Se_{0.5}$ sample was found to be a bulk superconductor. The $O_2$-annealed samples exhibited the coexistence of AFM order and weak superconductivity for $x \leq 0.1$. As the long-range AFM order was completely suppressed, the $O_2$-annealed samples with $x \geq 0.1$ became the bulk superconductors. The bottom line seems to be that the bulk superconducting region dramatically spreads by $O_2$-annealing. From these results, we concluded that $O_2$-annealing is the key factor in inducing bulk superconductivity for $FeTe_{1-x}Se_x$. To estimate the amount of oxygen, we carefully measured sample mass of $Fe_{1+x}Te_{0.9}Se_{0.1}$ before and after oxygen annealing. From the increase of the sample mass, the molecular formula of oxygen-annealed sample found to be $Fe_{1+x}Te_{0.9}Se_{0.1}O_{0.07}$. The excess Fe concentration $x$ have been measured to be $0.05 \lesssim x < 0.08$ as reported in Ref. 25 and 26. We found that the amount of intercalated oxygen between superconducting layers is comparable to that of excess Fe.

Neutron scattering measurements have revealed that the magnetic wave vectors $Q_d$ and $Q_s$ are observed over a wide composition range where $FeTe_{1-x}Se_x$ exhibits weak superconductivity [17, 27]. For our as-grown samples, weak superconductivity was observed in the range $0.1 \leq x \leq 0.4$, suggesting that the nesting vectors $Q_d$ and $Q_s$ coexist. It is expected that the nesting vector $Q_s$ becomes dominant for $FeTe_{0.5}Se_{0.5}$ where bulk superconductivity sets in. On the other hand, for the $O_2$-annealed samples, the bulk superconducting region extends down to $x = 0.1$, which implies that the nesting vector $Q_d$ is strongly suppressed by $O_2$-annealing. This suggests that the magnetic wave vector $Q_d$ is not favorable for superconducting paring [13, 27, 28]. According to first-principles density functional calculations [13], a small amount of excess Fe is responsible for the appearance of the magnetic wave vector $Q_d$ as the result of changing the Fermi surface that is driven by the electron doped from the excess Fe ion [28]. The oxygen ion could be intercalated between the layers and produces hole carriers which compensates the electron contributed by the excess Fe ion. Thus, the magnetic wave vector $Q_d$ could be strongly suppressed and the magnetic wave vector $Q_s$ becomes dominant and bulk superconductivity appears in the $O_2$-annealed samples.

4. Conclusion

We summarized the phase diagrams of as-grown and $O_2$-annealed $FeTe_{1-x}Se_x$. As-grown samples exhibit weak superconductivity in the region $0.1 \leq x \leq 0.4$ and bulk superconductivity for $x$

$\geq 0.5$.   On the other hand, the $O_2$-annealed samples show weak superconductivity in the range $0.025 \leq x < 0.1$ and bulk superconductivity for $x \geq 0.1$.   We found that $O_2$-annealing induces bulk superconductivity for FeTe$_{1-x}$Se$_x$.   This result implies that the magnetic wave vector $\mathbf{Q}_s$ is much stronger than the $\mathbf{Q}_d$ vector.   The oxygen ion intercalated between superconducting layers probably plays a key role in the suppression of the magnetic wave vector $\mathbf{Q}_d$ due to the compensation for the electron given by the excess Fe ion.

Acknowledgements
This work was partially supported by Grant-in-Aid for Scientific Research (KAKENHI).

References

[1] Y. Kamihara, T. Watanabe, M. Hirano and H. Hosono, J. Am. Chem. Soc. **130** (2008) 3296.

[2] M. Rotter, M. Tegel and D. Johrendt, Phys. Rev. Lett. **101** (2008) 107006.

[3] X. C. Wang, Q.Q. Liu, Y.X. Lv, W.B. Gao, L.X. Yang, R.C. Yu, F.Y. Li and C.Q. Jin, Solid State Commun. **148** (2008) 538.

[4] F. C. Hsu, J. Y. Luo, K. W. Yeh, T. K. Chen, T. W. Huang, P. M. Wu, Y. C. Lee, Y. L. Huang, Y. Y. Chu, D. C. Yan and M. K. Wu: Proc. Natl. Acad. Sci. U. S. A. **105** (2008) 14262.

[5] H. Haraldsen, F. Gronvold and J. Vihovde Tidsskrf. Kjemi Bergv. **4** (1944) 96.

[6] S. Li, C. de la Cruz, Q. Huang, Y. Chen, J. W. Lynn, J. Hu, Y. L. Huang, F. C. Hsu, K. W. Yeh, M. K. Wu and P. Dai, Phys. Rev. B **79** (2009) 054503.

[7] W. Bao, Y. Qiu, Q. Huang, M. A. Green, P. Zajdel, M. R. Fitzsimmons, M. Zhernenkov, S. Chang, M. Fang, B. Qian, E. K. Vehstedt, J. Yang, H. M. Pham, L. Spinu and Z. Q. Mao: Phys. Rev. Lett. **102** (2009) 247001.

[8] Y. Qiu. Wei Bao, Y. Zhao, C. Broholm, V. Stanev, Z. Tesanovic, Y. C. Gasparovic, S. Chang, J. Hu, B. Qian, M. Fang and Z. Mao, Phys. Rev. Lett. **103** (2009) 067008.

[9] H. Shi, Z. B. Huang, J. S. Tse, and H. Q. Lin: J. Appl. Phys. **110** (2011) 043917.

[10] M. A. McGire, A. D. Christianson, A. S. Sefat, B. C. Sales, M. D. Lumsden, R. Jin, E. A. Payzant and D. Mandrus, Phys. Rev. B **78** (2008) 094517.

[11] C. de la Cruz, Q. Huang, J. W. Lynn, J. Li, W. Ratcliff, J. L. Zarestky, H. A. Mook, G. F. Chen, J. L. Luo, N. L. Wang and P. Dai1, Nature **453** (2008) 899.

[12] Q. Huang, Y. Qiu, W. Bao, M. A. Green, J.W. Lynn, Y. C. Gasparovic, T. Wu, G. Wu and X. H. Chen, Phys. Rev. Lett. **101** (2009) 247001.

[13] M. J. Han and S. Y. Savrasov, Phys. Rev. Lett. **103** (2009) 067001.

[14] Y. Xia, D. Qian, L. Wray, D. Hsieh, G. F. Chen, J. L. Luo, N. L. Wang and M. Z. Hasan, Phys. Rev. Lett. **103** (2009) 037002.

[15] K. Nakayama, T. Sato, P. Richard, T. Kawahara, Y. Sekiba, T. Qian, G. F. Chen, J. L. Luo, N. L. Wang, H. Ding and T. Takahashi, Phys. Rev. Lett. **105** (2009) 197001.

[16] Y. Mizuguchi and Y. Takano, J. Phys. Soc. Jpn. **79** (2010) 102001.

[17] T. J. Liu J. Hu1, B. Qian, D. Fobes, Z. Q. Mao,W. Bao, M. Reehuis, S. A. J. Kimber, K. Prokes, S. Matas, D. N. Argyriou, A. Hiess, A. Rotaru, H. Pham, L. Spinu, Y. Qiu, V. Thampy, A. T. Savici, J. A. Rodriguez and C. Broholm, Nat. Mater. **9** (2010) 718.

[18] C. Dong, H. Wang, Z. Li, J. Chen, H. Q. Yuan and M. Fang, Phys. Rev. B **84** (2011) 224506.

[19] R. Khasanov, M. Bendele, A. Amato, P. Babkevich, A. T. Boothroyd, A. Cervellino, K. Conder, S. N. Gvasaliya, H. Keller, H. H. Klauss, H. Luetkens, V. Pomjakushin, E. Pomjakushina and B. Roessli, Phys. Rev. B **80** (2009) 140511.

[20] Y. Mizuguchi, F. Tomioka, S. Tsuda, T. Yamaguchi and Y. Takano, Appl. Phys. Lett. **94** (2009)

012503.

[21] K. Deguchi, Y. Mizuguchi, S. Ogawara, T. Watanabe, S. Tsuda, T. Yamaguchi and Y. Takano, Physica C **470** (2010) S340.

[22] Y. Mizuguchi, K. Deguchi, S. Tsuda, T. Yamaguchi and Y. Takano, Phys. Rev. B **81** (2010) 214510.

[23] Y. Mizuguchi, K. Deguchi, S. Tsuda, T. Yamaguchi and Y. Takano, Europhys. Lett. **90** (2010) 57002.

[24] Y. Mizuguchi, F. Tomioka, S. Tsuda, T. Yamaguchi and Y. Takano, J. Phys. Soc. Jpn. **78** (2009) 074712.

[25] R. Viennois, E. Giannini, D. van der Marel and R. Cerny, J. Solid State Chem. **183** (2010) 769.

[26] M. Onoda, Y. Kawasaki, M. Tsubokawa and T. Koyano, J. Phys.: Condens. Matter **22** (2010) 505702.

[27] M. D. Lumsden, A. D. Christianson, E. A. Goremychkin, S. E. Nagler, H. A. Mook, M. B. Stone, D. L. Abernathy, T. Guidi, G. J. MacDougall, C. de la Cruz, A. S. Sefat, M. A. McGuire, B. C. Sales and D. Mandrus, Nat. Phys. **6** (2010) 182.

[28] L. Zhang, D. J. Singh and M. H. Du, Phys. Rev. B **79** (2009) 012506.

Figure captions.

Fig. 1. Se concentration dependence of lattice constant $a$ and $c$ for the as-grown and $O_2$-annealed $FeTe_{1-x}Se_x$ single crystals.

Fig. 2. Temperature dependence of magnetic susceptibility for $FeTe_{1-x}Se_x$ (a) $x = 0.1$ and (b) $x = 0.4$ under several $O_2$-annealing conditions.

Fig. 3. Temperature dependence of normalized magnetic susceptibility around $T_N$ for (a) as-grown and (c) $O_2$-annealed $FeTe_{1-x}Se_x$. The arrows indicate the magnetic transition temperature. Temperature dependence of magnetic susceptibility around $T_c$ for (b) as-grown and (d) $O_2$-annealed $FeTe_{1-x}Se_x$.

Fig. 4. Se concentration dependence of shielding volume fraction for as-grown and $O_2$-annealed $FeTe_{1-x}Se_x$.

Fig. 5. Phase diagrams showing $T_c$ and $T_N$ as a function of $x$ for (a) as-grown $FeTe_{1-x}Se_x$ and (b) $O_2$-annealed $FeTe_{1-x}Se_x$.

Fig. 1.

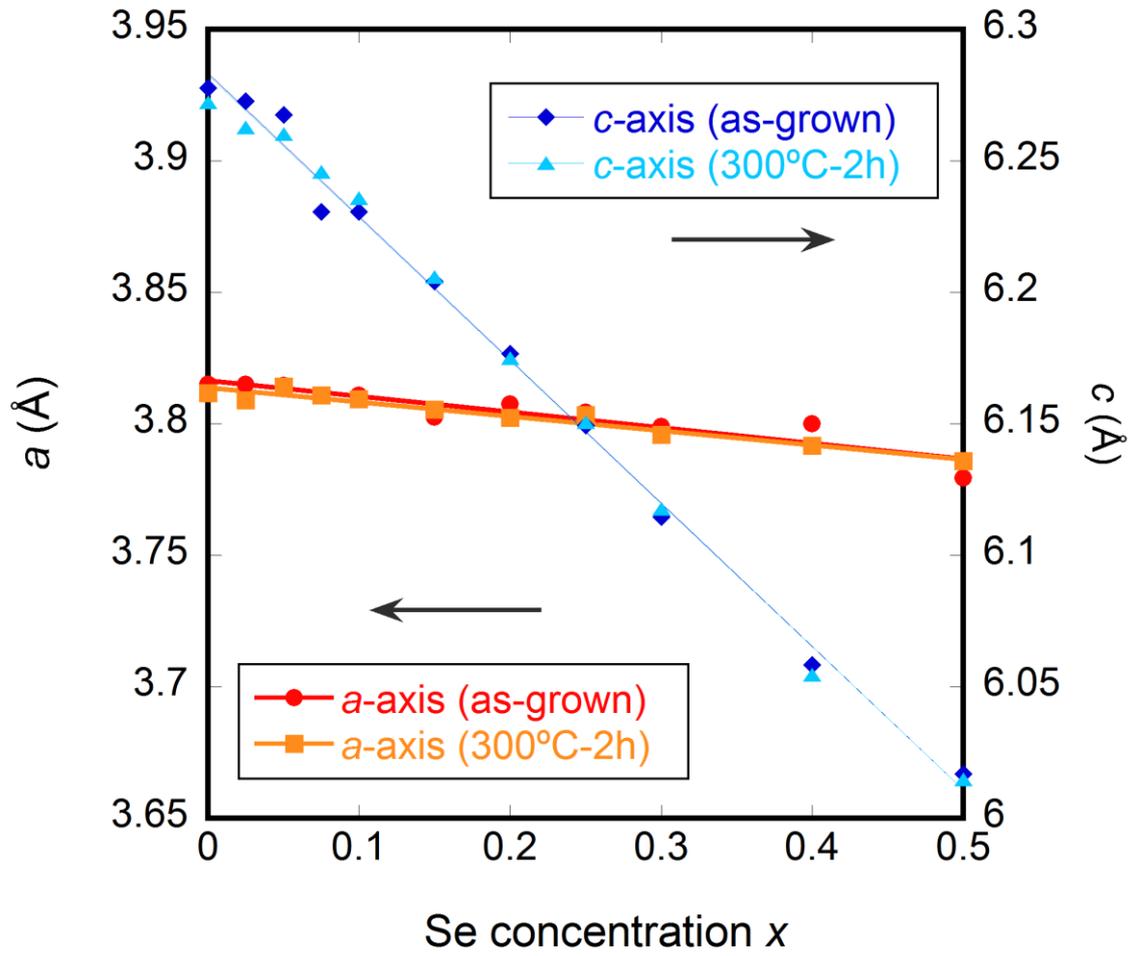

Fig. 2.

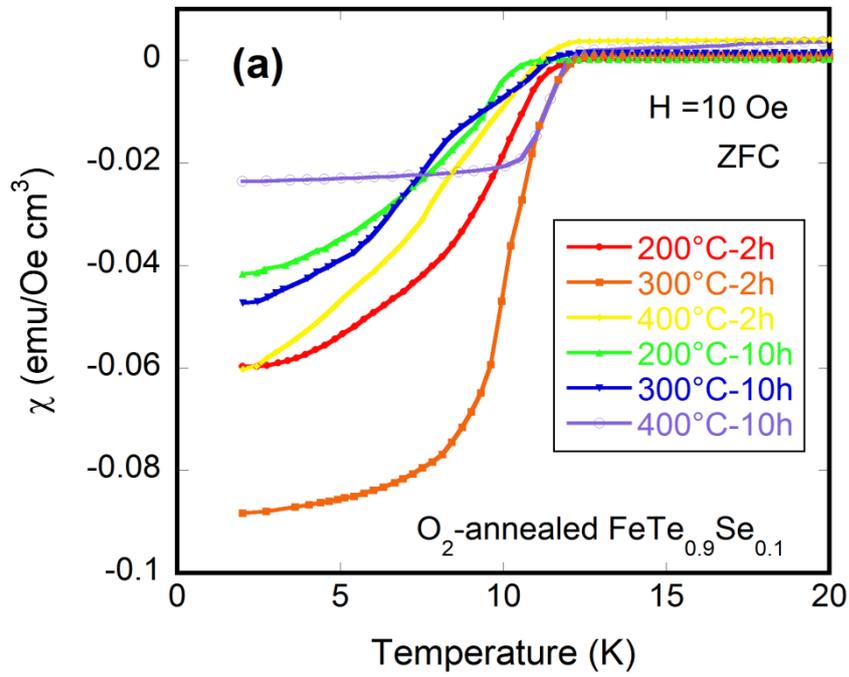

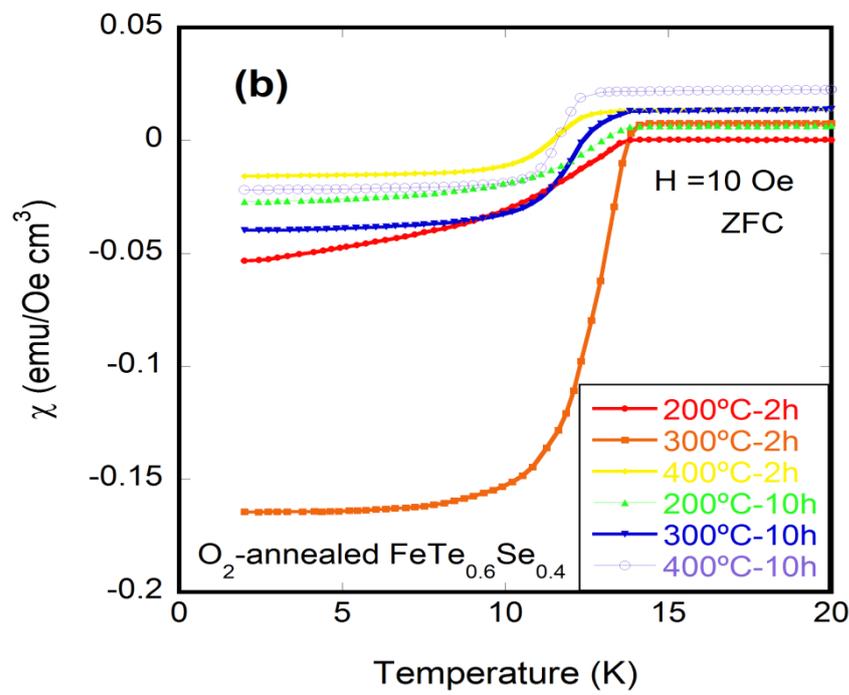

Fig. 3.

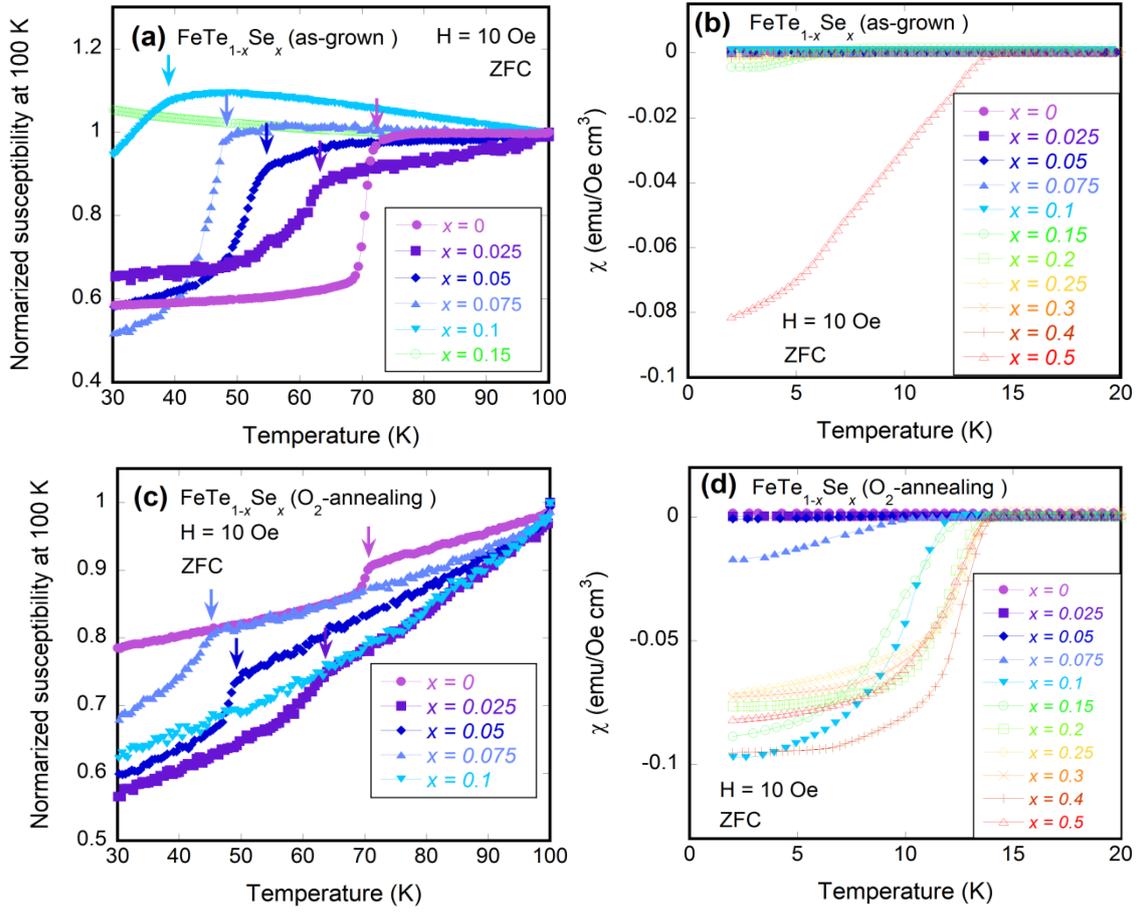

Fig. 4.

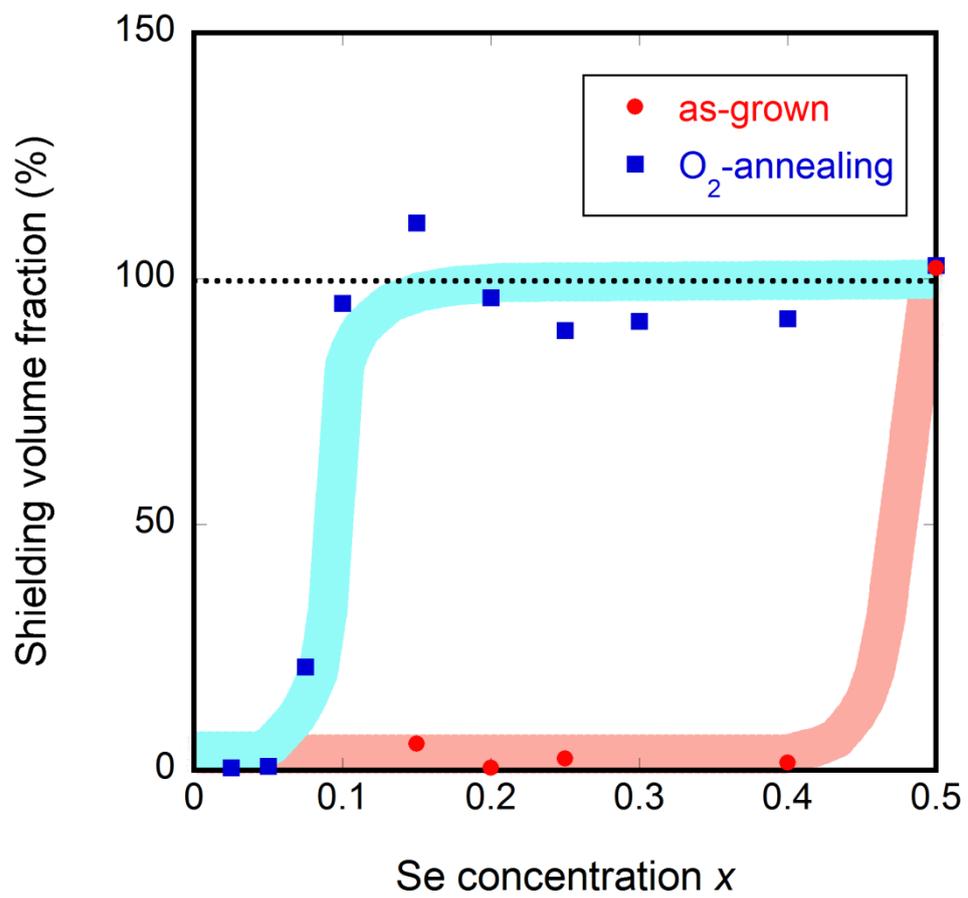

Fig. 5.

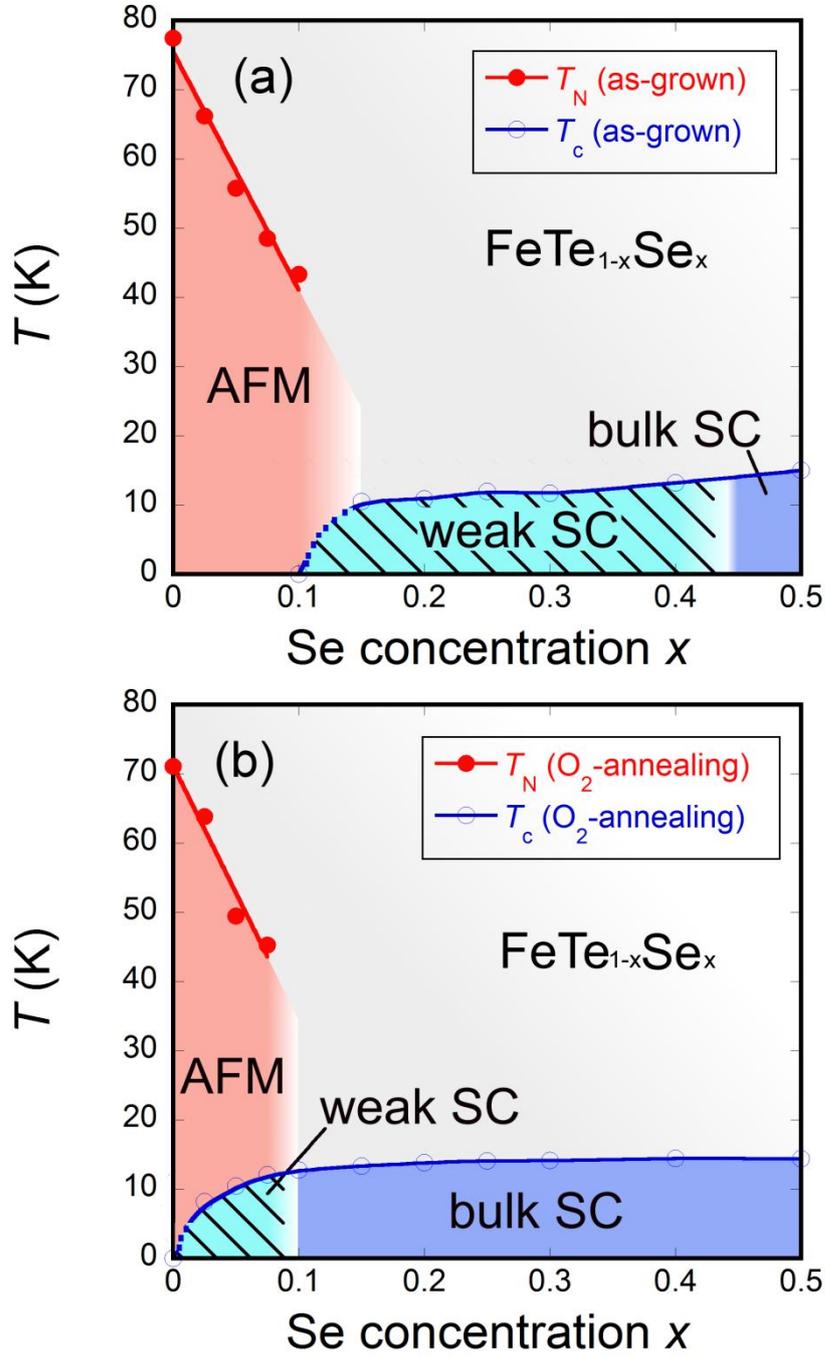